\documentclass[sn-mathphys]{sn-jnl}
\jyear{2021}%
\usepackage{subfigure}
\usepackage{tensor}

\theoremstyle{thmstyleone}%

\theoremstyle{thmstyletwo}%

\theoremstyle{thmstylethree}%

\begin{document}

\title[ ]{Analytical Stellar Models of Neutron Stars in Teleparallel Gravity}

\author[1]{\fnm{Jay} \sur{Solanki}}\email{jay565109@gmail.com}

\author[2]{\fnm{Rohan} \sur{Joshi}}\email{joshirohan043@gmail.com}

\author[2]{\fnm{Malay} \sur{Garg}}\email{gargmalay76@gmail.com}

\affil[1]{\orgname{Sardar Vallabhbhai National Institute of Technology}, \orgaddress{\city{Surat}, \postcode{395007}, \state{Gujarat}, \country{India}}}

\affil[2]{\orgname{Delhi Technological University}, \orgaddress{\city{Delhi}, \postcode{110042}, \country{India}}}

\abstract{In this paper, we developed three analytical models and obtained a new class of solutions describing compact stellar structures using the theory of teleparallel gravity. We consider the general anisotropic nature of stellar configurations and solve teleparallel gravity equations. In order to thoroughly analyze the various parameters of the stars, we developed three models by choosing various physically acceptable forms of metric potential $ e^{d(r)} $ and radial pressure $ p_r(r) $. We also analyze the impact of teleparallel gravity's parameters $ \beta $ and $ \beta_1 $ on the description of the stellar structures. We calculated model parameters such that models describing various observed neutron stars obey all physical conditions to be potentially stable and causal. By analyzing the impact of various parameters of teleparallel gravity on the description of anisotropic stellar structures, we found that three models developed in this paper can describe anisotropic neutron stars ranging from low density to high density. Finally, we obtain a quadratic Equation of State for each model describing various neutron stars, which can be utilized to find compositions of the stellar structures. It is very useful to find models that can exhibit quadratic EOS, since material compositions of real neutron stars and strange stars are found to exhibit quadratic EOS by various authors. Non linear $ f(T) $ model gives high deviation of EOS from quadratic behaviour, thus, in this paper we work with linear $ f(T) $ function by using diagonal tetrad to model realistic compact stars.}




\maketitle

\section{Introduction}\label{Sec1}
It is well known that general relativity describes the effect of gravity as an effect of the curvature of space-time. Thus, general relativity is a geometrical theory of gravitation. General relativity is a very successful theory describing various astronomical and astrophysical phenomena with great precision. However, recent observations of galactic dynamics can not be understood in the tenets of general relativity. Thus, the concepts of dark matter and dark energy have been introduced to understand the phenomena of galactic dynamics and the expansion of the universe.\cite{COPELAND_2006, sami2009dark, sami2009primer, 2011AdAst2011E...8G} However, another way to account for observations of galactic dynamics like phenomena that can not be explained by general relativity is to modify general relativity.\cite{RevModPhys.82.451, durrer2008dark, PhysRevD.70.043528, PhysRevD.81.127301, Capozziello_2011, Nojiri:2010wj, Nojiri:2017ncd, Nojiri:2003ft} The Lagrangian formulation of the general relativity is constructed from Einstein-Hilbert action written in terms of Ricci scalar and metric tensor.\cite{RevModPhys.82.451} The modified gravity can be constructed by modifying the Einstein-Hilbert action. Modified field equations can be obtained by varying modified Einstein-Hilbert action and implementing the Euler-Lagrangian equation. In recent years there is a huge effort to modify gravity to solve the issue of non-renormalizability of general relativity.\cite{PhysRevLett.108.031101, PhysRevD.16.953}
\par The general relativity is constructed from its dynamical variable - the metric tensor, defined on a pseudo-Riemannian manifold. In general relativity, connections are considered to be torsion-less. However, in order to account for torsion, Einstein introduced  Teleparallel Equivalent of General Relativity known as TEGR.\cite{1985FoPh...15..365I, PhysRevD.19.3524} In this theory, Einstein introduced absolute parallelism and the tangent space as a frame of reference to understand the effect of torsion. In TEGR, tetrad fields are used instead of metric tensor to construct the theory.\cite{PhysRevD.19.3524} Because of using tangent space frame of reference, the curvature at each point becomes zero, and the only effect that remains is of torsion.\cite{doi:10.1142/S0218271808013972} Thus, TEGR describes the gravity as due to the effect of torsion of space-time. TEGR also can be constructed by Lagrangian formulation by considering tetrad field as dynamical variables instead of the metric tensor. Thus, we can also modify this action to construct modified gravity in terms of the torsion mechanism. This modified theory of gravity is known as $ f(T) $ gravity.\cite{Cai_2016, PhysRevD.79.124019, Kr_k_2016}. Also, recently the reference \cite{bahamonde2021teleparallel} has provided a detailed review on the topic of teleparallel gravity and modified $ f(T) $ gravity, which readers can consider for more discussion on the subject. 
\par In recent years, teleparallel gravity is found to be very successful in describing various astronomical and astrophysical phenomena. $ f(T) $ gravity is also a widely used theory besides $ f(R) $ gravity to explain the expansion of the universe and dark matter through modification of general relativity.\cite{PhysRevD.84.083518, Momeni_2015, Momeni_2014a, Jamil_2014, Daouda_2011, El_Hanafy_2016a, Jamil_2012a, RODRIGUES_2013, HOUNDJO_2012, Jamil_2012b, El_Hanafy_2016b, Jamil_2012c, Jamil_2012d, Jamil_2013, Jamil_2012e, Jamil_2012f, Farooq_2013} The compact astrophysical objects such as black hole and compact stars have been modeled using TEGR\cite{B_hmer_2011, Ulhoa_2013, article1, Zubair_2015,  Abbas_2015, article2, article3, article4} as well as other theories of modified gravity\cite{Astashenok:2014nua, Capozziello:2015yza, Astashenok:2020qds, Astashenok:2021peo, Astashenok:2020cqq}. Recently Wang et al. studied spherically symmetric static solution in $f(T)$ gravity models \cite{PhysRevD.84.024042}. It is found that only a limited class of $f(T)$ gravity models can be solved in this frame. Cesmsinnan Deliduman and Baris Yapiskan investigated neutron stars under modified $f(T)$ gravity. They found that the relativistic neutron star solution in $f(T)$ gravity models is possible only if $f(T)$ is a linear function of the torsion scalar $T$ for diagonal tetrad. The black hole with a cosmological constant has been studied in this paper\cite{Ferraro_2011}.  The violation of the first law of black hole thermodynamics in $f(T)$ gravity due to the lack of local Lorentz invariance is studied in this work \cite{miao2011violation}. 
\par Thus, we choose linear function $ f(T) $, to be $ f(T) = \beta T + \beta_1 $. This is essentially known as teleparallel gravity with the extra factors $ \beta $ and $ \beta_1 $.  Many authors studied the effect of this linear $ f(T) $ on the modeling of anisotropic neutron stars. However, many of the studies are performed by setting the parameter $ \beta_1 $ equals zero to describe and analyze graphs of various parameters of neutron stars.\cite{Abbas_2015, article2, article4} In this paper, we developed three models to understand the role of various parameters like metric potential, radial pressure, $ \beta $, $ \beta_1 $, etc. on the properties of the neutron stars. We considered different physically acceptable forms of metric potential $ e^{d(r)} $ and radial pressure $ p_r(r) $, to solve equations of teleparallel gravity and to generate models that can describe various types of neutron stars ranging from low density to high density. In first two models we considered $\beta$ to be a constant of value $ \beta = 2 $. Then we compute the physically acceptable value of $ \beta_1 $ along with other parameters of the models. Thus, in the first two models, we analyze the effect of parameter $ \beta_1 $ on the properties of observed neutron stars. In third model, we considered $ \beta_1 = 0 $, and observed the effect of parameter $ \beta $ on the properties of observed neutron stars. We found that these three models can describe anisotropic neutron stars ranging from low density to high density.
\par We organized this paper as follows: In section (\ref{Sec2}), we formulate the field equations of teleparallel gravity. In that section, we briefly review the idea of the tetrad field and write down the elements of the manifold in terms of tetrad field. By varying the action and considering the Weitzenb\"{o}ck connection, we write down equations of teleparallel gravity for anisotropic matter distribution. In sections (\ref{Sec3}), (\ref{Sec4}), and (\ref{Sec5}), we developed the first, second, and third model describing anisotropic neutron stars in teleparallel gravity. In each of those sections we developed the model by considering physically acceptable forms of metric potential $ e^{d(r)} $ and radial pressure $ p_r(r) $. After that, we impose physical conditions so that model becomes physically acceptable. After that, we calculated various model parameters for describing observed neutron stars. We calculated model parameters and neutron star's physical parameters for each neutron star mentioned in this paper. We also plotted graphs of various parameters of neutron stars to study the general behavior of each model on the description of the stellar structures. Finally, we obtain a quadratic Equation of State for each model describing various neutron stars, which can be utilized to find compositions of the stellar structures. Finally, we summarize the results obtained through discussion section (\ref{Sec6}).

\section{Formulation of the Field Equations in the Teleparallel Gravity}\label{Sec2}
General relativity is constructed from its dynamical variable - the metric tensor, defined on a pseudo-Riemannian manifold. However, in teleparallel gravity tetrad field is used as a dynamical variable, which forms an orthonormal basis in tangent space at each point. Now let us denote Latin indices ($i$, $j$, ...) for coordinates of tangent space and greek indices ($\mu$, $\nu$, ...) for space-time coordinates. Thus basis vectors and covectors for tangent space are given by $ e_i = \partial_i $ and $ e^i = dx^i $. Basis vectors for space-time coordinates are given by $ e_\mu = \partial_\mu $ and $ e^\mu = dx^\mu $. The vectors and covectors can be transformed from one base to another as follow
\begin{equation}
\label{1}
    \partial_i = e\indices{_i^\mu}\partial_\mu \quad\mathrm{and}\quad dx^i = e\indices{^i_\mu} dx^\mu
\end{equation}    
and conversely
\begin{equation}
\label{2}
    \partial_\mu = e\indices{^i_\mu}\partial_i \quad\mathrm{and}\quad dx^\mu = e\indices{_i^\mu} dx^i
\end{equation}    
where $e\indices{^i_\mu}$ is known as tetrad field and $ e\indices{_i^\mu} $ is inverse of it, that satisfy the relations $ e\indices{^i_\mu}e\indices{_j^\mu} = \delta^i_j $ and $e\indices{^i_\mu}e\indices{_i^\nu} = \delta^\nu_\mu $. The relation between the metric of space-time and the tetrad field is given by
\begin{equation}
\label{3}
    g_{\mu \nu} = \eta_{ij} e\indices{^i_\mu}e\indices{^j_\nu}
\end{equation}
where $ \eta_{i j} = diag[1, -1, -1, -1] $. From equation (\ref{1}), line element of the manifold is given in terms of tetrad field as
\begin{equation}
\label{4}
    ds^2 = g_{\mu\nu}dx^\mu dx^\nu = \eta_{ij} e\indices{^i_\mu}e\indices{^j_\nu} dx^\mu dx^\nu
\end{equation}
Now in teleparallel gravity, the curvature-less connection known as Weitzenb\"{o}ck connection is given by 
\begin{equation}
\label{5}
    \Gamma\indices{^\alpha_\mu_\nu} =  e\indices{_i^\alpha} \partial_\nu e\indices{^i_\mu}
\end{equation}
Now the torsion tensor in terms of Weitzenb\"{o}ck connection is given by
\begin{equation}
\label{6}
    T\indices{^\alpha_\mu_\nu} = \Gamma\indices{^\alpha_\nu_\mu} - \Gamma\indices{^\alpha_\mu_\nu}
\end{equation}
Now con-torsion and superpotential tensors have been introduced to calculate torsion scalar. The con-torsion and superpotential tensors are respectively
\begin{equation}
\label{7}
     K\indices{^\mu^\nu_\alpha} =  -\frac{1}{2}(T\indices{^\mu^\nu_\alpha} -  T\indices{^\nu^\mu_\alpha} -  T\indices{_\alpha^\mu^\nu})
\end{equation}
\begin{equation}
\label{8}
     S\indices{_\alpha^\mu^\nu} =  \frac{1}{2}(K\indices{^\mu^\nu_\alpha} + \delta^\mu_\alpha T\indices{^\rho^\nu_\rho} - \delta^\nu_\alpha T\indices{^\rho^\mu_\rho})
\end{equation}
Now the torsion scalar can be defined as
\begin{equation}
\label{9}
    T = S\indices{_\alpha^\mu^\nu}T\indices{^\alpha_\mu_\nu}
\end{equation}
Now in $ f(T) $ gravity, modified gravitational action is written in terms of torsion scalar in the units of (G = c = 1) as
\begin{equation}
\label{10}
    S = \int d^4x\left[ \frac{1}{16\pi}f(T) + \mathcal{L}_m \right]e
\end{equation}
Where $f(T)$ is a function of torsion scalar and $\mathcal{L}_m$ is Lagrangian density for matter field. Also $ e = \sqrt{-g} = det[e^i_\mu] $.
\\ Now the variation in action (10) gives the field equations of $ f(T) $ gravity. \cite{article2, Jamil_2013, B_hmer_2012}
\begin{equation}
\label{11}
    S\indices{_\mu^\nu^\gamma} \partial_\gamma T f_{TT} + e^{-1} e\indices{^i_\mu} \partial_\gamma(e e\indices{_i^\alpha} S\indices{_\alpha^\nu^\gamma})f_T + T\indices{^\alpha_\rho_\mu}S\indices{_\alpha^\nu^\rho}f_T + \frac{1}{4}\delta^\nu_\mu f = 4\pi \mathcal{T}\indices{^\nu_\mu}
\end{equation}
Where $\mathcal{T}_{\mu\nu}$ is the energy- momentum tensor of matter and $f_T$ and $ f_{TT} $ are first and second derivative of $ f(T) $ with respect to T respectively. Now in the case of anisotropic matter distribution, the energy-momentum tensor is given by
\begin{equation}
\label{12}
    \mathcal{T}\indices{^\nu_\mu} = (\rho + p_t)u_\mu u^\nu - p_t\delta^\nu_\mu + (p_r - p_t)v_\mu v^\nu
\end{equation}
Where $ \rho $, $ p_r $ and $ p_t $ are energy density, radial pressure and tangential pressure, respectively. $ u^\mu $ and $ v^\mu $ are the four velocity and radial four vectors, respectively. We define anisotropy of the compact stellar structure as $ S = \frac{p_r - p_t}{\sqrt{3}} $.

\section{First Anisotropic Model of Neutron Star in Teleparallel Gravity}\label{Sec3}
We consider the static and spherically symmetric model of the star. Thus space-time metric describing interior of the star is given by 
\begin{equation}
\label{13}
ds^2 = e^{c(r)}dt^2 - e^{d(r)}dr^2 - r^2(d\theta^2+sin^2\theta d\phi^2)
\end{equation}
For the metric (\ref{13}), we calculate the tetrad matrix as
\begin{equation}
\label{14}
    e^i_\mu = diag[e^{\frac{c(r)}{2}}, e^{\frac{d(r)}{2}}, r, r sin(\theta)]
\end{equation}
From equation (\ref{14}) we obtain,
\begin{equation}
\label{15}
    e = det[e^i_\mu] = e^{\frac{(c + d)}{2}}r^2sin(\theta)
\end{equation}
Now we determine torsion scalar as 
\begin{equation}
\label{16}
    T(r) = \frac{2 e^{-d(r)}}{r}\left( c' + \frac{1}{r} \right)
\end{equation}
Now from equation (\ref{11}), we write equation of $f(T)$ gravity for the metric (\ref{13}) as \cite{article2, Jamil_2013, B_hmer_2012}
\begin{equation}
\label{17}
  4\pi\rho =  \frac{f}{4} - \frac{f_T}{2}\left( T - \frac{e^{-d}}{r}(c' + d') - \frac{1}{r^2} \right)
\end{equation}

\begin{equation}
\label{18}
    4\pi p_r = \frac{f_T}{2}\left( T - \frac{1}{r^2} \right) - \frac{1}{4}
\end{equation}

\begin{equation}
\label{19}
    4\pi p_t = \left[ \frac{T}{2} + e^{-d}\left( \frac{c''}{2} + \left( \frac{c'}{4} + \frac{1}{2r} \right)\left( c' - d' \right) \right) \right]\frac{f_T}{2} - \frac{f}{4}
\end{equation}

\begin{equation}
\label{20}
    \frac{cot(\theta)}{2r^2}T'f_{TT} = 0
\end{equation}

where single prime and double prime denotes first and second derivative with respect to r. Now equation (\ref{20}) leads to the following linear form of $ f(T) $
\begin{equation}
\label{21}
    f(T) = \beta T + \beta_1
\end{equation}
where $\beta$ and $\beta_1$ are constants. Thus, the theory becomes comparable to teleparallel gravity with the extra factors of $ \beta $ and $ \beta_1 $ and we will study the effects of these parameters by generating different three models. Now in order to solve equations (\ref{17}) to (\ref{19}), we choose following form of metric $e^{d(r)}$,
\begin{equation}
\label{22}
    e^{d(r)} = 1 - \frac{a r^2}{R^2}
\end{equation}
where a is constant and R is the radius of the neutron star. We consider $ a < 1 $, so that the metric does not become singular at any point for $ r < R $. In this model we consider $ \beta = 2 $ in equation (\ref{21}). By substituting equation (\ref{22}) in equation (\ref{17}), we obtain 
\begin{equation}
\label{23}
    4\pi \rho = \frac{-\frac{a}{R^2}\left( 3 - \frac{a r^2}{R^2} \right)}{\left( 1 - \frac{a r^2}{R^2} \right)^2} + \frac{\beta_1}{4}
\end{equation}
Now from the equation (\ref{18}), we write
\begin{equation}
\label{24}
    c'(r) = (4\pi p_r)re^{d(r)} + \frac{e^{d(r)} - 1}{r} + \frac{\beta_1 r e^{d(r)}}{4}
\end{equation}
In order to solve equation (\ref{24}), we consider following form of $ 4\pi p_r $
\begin{equation}
\label{25}
    4\pi p_r = \frac{\frac{p_0}{R^2}\left( 1 - \frac{r^2}{R^2} \right)}{1 - \frac{a r^2}{R^2}}
\end{equation}
where $ p_0 $ is another constant, that is the parameter of the model such that $ \frac{p_0}{R^2} $ denotes the central pressure of the neutron star. The constraint on the $ p_0 $ is that $ p_0 > 1 $, so that radial pressure remains positive for $ 0 < r < R $. The particular form of $ p_r $ is physically reasonable because it is monotonically decreasing function of r and satisfy the condition of vanishing radial pressure at the surface of the neutron star $ r = R $. by substituting values of equations (\ref{22}) and (\ref{25}) into (\ref{24}), we obtain
\begin{equation}
\label{26}
    c'(r) = r\left(\frac{p_0 - a}{R^2} + \frac{\beta_1}{4}\right) - r^3\left( \frac{p_0}{R^4} + \frac{\beta_1 a}{4 R^2} \right)
\end{equation}
by integrating above equation, we obtain the metric as
\begin{equation}
\label{27}
    e^{c(r)} = K e^{ \frac{r^2}{2}\left(\frac{p_0 - a}{R^2} + \frac{\beta_1}{4}\right) - \frac{r^4}{4}\left( \frac{p_0}{R^4} + \frac{\beta_1 a}{4 R^2} \right)}
\end{equation}
where $ K $ is the integration constant. Now interior space-time of the neutron star in this model is given by
\begin{equation}
\label{28}
\begin{aligned}
    ds^2 = {}& \left( K e^{ \frac{r^2}{2}\left(\frac{p_0 - a}{R^2} + \frac{\beta_1}{4}\right) - \frac{r^4}{4}\left( \frac{p_0}{R^4} + \frac{\beta_1 a}{4 R^2} \right)} \right)dt^2 - \left( 1 - \frac{a r^2}{R^2} \right)dr^2 \\ &- r^2(d\theta^2+sin^2\theta d\phi^2)
\end{aligned}    
\end{equation}
\subsection{Constraining parameters for physically well-behaved model}
\subsubsection{Matching exterior space-time with interior space-time}
Now the space-time metric should be continuous at the boundary of the star. Thus, interior metric (\ref{28}) of the star should be matched to exterior Schwarzschild space-time metric at $ r = R $ in this model.
\begin{equation}
\label{29}
ds^2 = \left(1 - \frac{2M}{r}\right)dt^2 - \left(1 - \frac{2M}{r}\right)^{-1}dr^2 - r^2(d\theta^2 + sin^2\theta d\phi^2)
\end{equation}
Now interior metric (\ref{28}) will be matched to exterior Schwarzschild metric at the surface of the star if 
\begin{equation}
\label{30}
    1 - a = \left(1 - \frac{2M}{R}\right)^{-1}
\end{equation}
and
\begin{equation}
\label{31}
    K e^{ \left(\frac{p_0 - 2a}{4} + \frac{\beta_1 R^2}{8}\left( 1 - \frac{a}{2}\right)\right)} =  1 - \frac{2M}{R}
\end{equation}
From equation (\ref{30}) and (\ref{31}) we obtain the value of $K$ as
\begin{equation}
\label{32}
    K = \frac{1}{1 - a} e^{- \left(\frac{p_0 - 2a}{4} + \frac{\beta_1 R^2}{8}\left( 1 - \frac{a}{2}\right)\right)}
\end{equation}
For given value of $ a $, $ p_0 $, $ \beta_1 $ and $ R $, value of constant $ K $ in equation (28) can be calculated from equation (\ref{32}). 
\subsubsection{Imposing physical conditions to the model}
Since compact stars are physical objects that exist in nature, their physical parameters, such as energy density and pressures, must obey the basic laws of physics incorporated in physical conditions. Thus, any physically acceptable model of the compact star should obey the following physical conditions for $ 0 < r < R $
\\ 
(i) \ $ \rho, p_r, p_t \ge 0 $ \\
(ii) \ $ \frac{d \rho}{dr}, \frac{d p_r}{dr}, \frac{d p_t}{dr} \le 0 $ \\
(iii) \ $ 0 \le \frac{d p_r}{d\rho}, \frac{d p_t}{d\rho} \le 1 $ \\
(iv) \ $ \frac{d p_t}{d\rho} - \frac{d p_r}{d\rho} < 0 $ \\
where condition (i) is known as weak energy condition, condition (ii) implies that the density and pressure of the star must be a decreasing function of radius. The condition (iii) is known as causality condition, which implies that radial and transverse speed of sound, $ v_r = \sqrt{\frac{d p_r}{d\rho}}  $ and $ v_t = \sqrt{\frac{d p_t}{d\rho}}  $ must not exceed the speed of light. The condition (iv) is known as the stability condition, which implies that the radial speed of sound must be greater than the transverse speed of sound for the star to be potentially stable for $ 0 < r < R $.\\
We have already chosen parameters such that $ a < 1 $ and $ p_0 > 0 $ for $ \rho, p_r \ge 0 $. We can get $ p_t \ge 0  $ if the following condition satisfies
\begin{multline}
     -64 p_0 + 64 a p_0 + R^4 \beta_1^2 + 3a^2(4 + R^2\beta_1)^2 \\- 
  a^3(4 + R^2\beta_1)^2 - a R^2 \beta_1(16 + 3 R^2 \beta_1) \ge 0
\end{multline}
Now in this model we already have $ \frac{d \rho}{dr}, \frac{d p_r}{dr} \le 0 $. In order to get $ \frac{d p_t}{dr} \le 0 $ we impose the following condition
\begin{multline}
    -R^4\beta_1^2 -2a^4R^2\beta_1(4 + R^2\beta_1) - a^3(4 + 8p_0 - 7R^2\beta_1)(4 + R^2\beta_1) + \\8p_0(16 + R^2\beta_1) +  a^2(-48 - 32R^2\beta_1 - 9R^4\beta_1^2 + 24p_0(4 + R^2\beta_1)) +\\ a(-24p_0(8 + R^2\beta_1) + R^2\beta_1(16 + 5R^2\beta_1)) \ge 0
\end{multline} 
Now conditions (iii) and (iv) satisfy if the model parameters satisfy following conditions, which we deduce by calculating $\frac{d p_r}{d\rho} $, $ \frac{d p_t}{d\rho} $ and $\frac{d p_t}{d\rho} - \frac{d p_r}{d\rho} $ at $ r = 0 $ and $ r = R $.
\begin{equation}
\label{35}
    0 \le \frac{(1 - a)p_0}{5a^2} \le 1 \quad\mathrm{and}\quad 0 \le \frac{(1 - a)^2p_0}{(5 - a)a^2} \le 1
\end{equation}
\begin{multline}
\label{36}
    0 \le \frac{-48a^2 + a(64p_0 - 16R^2\beta_1) + 16p_0^2* + R^4\beta_1^2 + 8p_0(-16 + R^2\beta_1)}{320a^2} \le 1
\end{multline}
\begin{multline}
\label{37}
    0 \le \frac{1}{64(-5 + a)a^2)}(R^4\beta_1^2 + 2a^4R^2\beta_1(4 + R^2\beta_1) + a^3(4 + 8p_0 - 7R^2\beta_1)(4 + R^2\beta_1)\\ - 8p_0(16 + R^2\beta_1) + a^2(48 + 32R^2\beta_1 + 9R^4\beta_1^2 - 24p_0(4 + R^2\beta_1))\\ + a(24p_0(8 + R^2\beta_1) -  R^2\beta_1(16 + 5R^2\beta_1))) \le 1
\end{multline}
\begin{equation}
\label{38}
     48a^2 + 16p_0^2 - 16a R^2\beta_1 + R^4\beta_1^2 + 
      8p_0(-8 + R^2\beta_1) > 0
\end{equation}
\begin{multline}
\label{39}
    R^4\beta_1^2 + 2a^4R^2\beta_1(4 + R^2\beta_1) + a^3(4 + 8p_0 - 7R^2\beta_1)(4 + R^2\beta_1) \\- 8p_0(8 + R^2\beta_1) +  a^2(48 + 32R^2\beta_1 + 9R^4\beta_1^2 - 8p_0(4 + 3R^2\beta_1)) \\+ a(8p_0(8 + 3R^2\beta_1) - R^2\beta_1(16 + 5R^2\beta_1)) > 0
\end{multline}

\subsection{Physical analysis}
In this section, we analyze the behavior of physical parameters such as energy density and two pressures at the interior of the star for five observed neutron stars. We calculate parameters for each star such that this first model describing five neutron stars mentioned in Table 1. satisfy all bound conditions from (i) to (iv).  We obtain Mass(M) and radius(R) of the neutron stars from references\cite{article4, Steiner_2010, 2011ApJ...730...25R} and compute central density($  \rho_c $), surface density($ \rho_s $), central pressure($p_r(r=0) = p_t(r=0) = p_c$) and surface tangential pressure ($ p_t(r=R) = p_s $) of each star, shown in Table 2. We have plotted various parameters of the star shown in figure \ref{fig1}. (a) to (h) of a neutron star SAX J1808.4-3658(SS2) (5) shown in Table 1. described by the first model of the neutron stars in teleparallel gravity. The figures indicate that all the physical parameters of the star are well-behaved and follow all conditions in equations (i) to (iv) at all interior points of the star.

\begin{table}[h]
\begin{center}
\caption{Values of the model parameters of the first model describing five observed neutron stars with $ \beta = 2 $}\label{tab1}%
\begin{tabular}{@{}llllll@{}}
\toprule
Star & R & M & a & $ \beta_1 \times 10^4 $ & $p_0$ \\ 
        & ($km$) & ($ M_{\odot} $) & & ($ km^{-2} $) \\
\midrule
 HER X-1 & 7.7 & 0.88 & -0.512 & -0.843 & 0.09  \\ 
 
 RX J1856-3754 & 12.7 & 1.4 & -0.486 & -0.248 & 0.08 \\

 SMC X-1 & 8.301 & 1.04 & -0.592 & -1.161 & 0.11  \\
 
 RX J1856.5-3754 & 12 & 1.4 & -0.529 & -0.382 & 0.095 \\
 
 SAX J1808.4-3658(SS2) & 7.951 & 0.9 & -0.506 & -0.759 & 0.089 \\
\botrule
\end{tabular}
\end{center}
\end{table}

\begin{table}[h]
\begin{center}
\caption{Values of the physical parameters of the five observed neutron stars described by the first model}\label{tab2}%
\begin{tabular}{@{}lllllll@{}}
\toprule
 Star & R & M & $ \rho_c \times 10^3 $ & $ \rho_s \times 10^3 $ & $p_c\times 10^4$ & $p_s\times 10^4$ \\ 
        & (km) & ($ M_{\odot} $) & ($ km^{-2} $) & ($ km^{-2} $) & ($ km^{-2} $) & ($ km^{-2} $)\\
\midrule
  HER X-1 & 7.7 & 0.88 & 2.06 & 1.05 & 1.21 & 0.55 \\ 
 
 RX J1856-3754 & 12.7 & 1.4 & 0.72 & 0.38 & 0.39 & 0.19\\

 SMC X-1 & 8.301 & 1.04 & 2.05 & 0.97 & 1.27 & 0.63\\
 
 RX J1856.5-3754 & 12 & 1.4 & 0.88 & 0.44 & 0.53 & 0.23\\
 
 SAX J1808.4-3658(SS2) & 7.951 & 0.9 & 1.91 & 0.98 & 1.12 & 0.49\\
\botrule
\end{tabular}
\end{center}
\end{table}

\subsection{Equation of state}
\begin{figure*}[ht!]
\centering
\subfigure[]{\includegraphics[width = .43\textwidth]{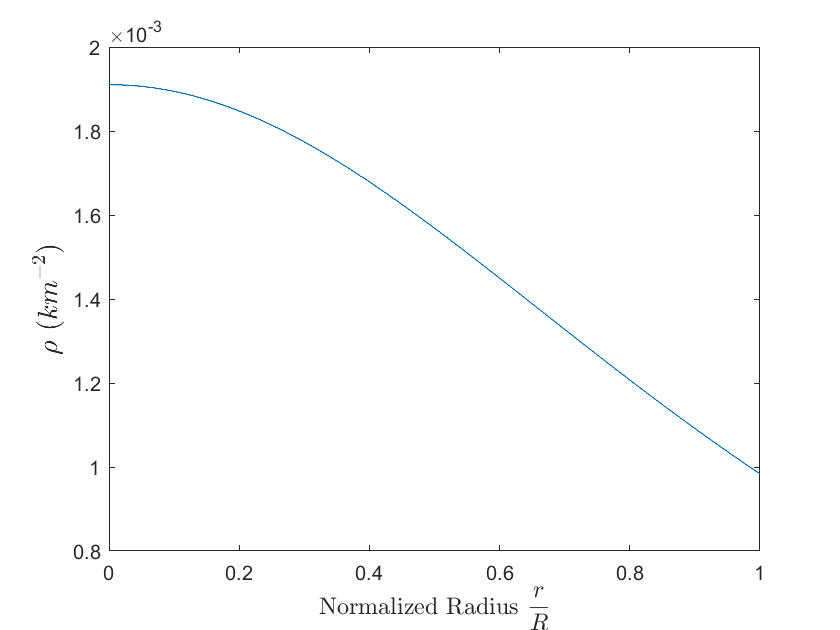}}
\subfigure[]{\includegraphics[width = .43\textwidth]{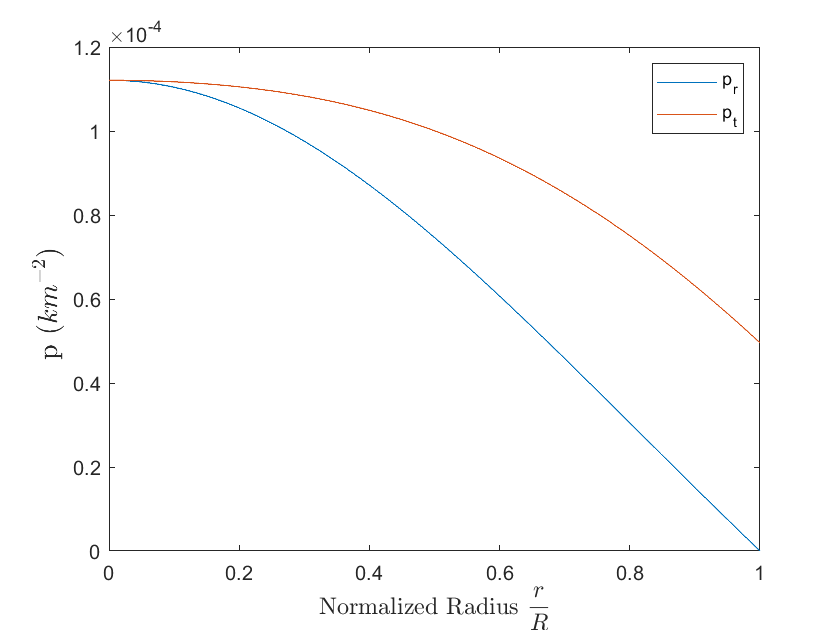}}
\subfigure[]{\includegraphics[width = .43\textwidth]{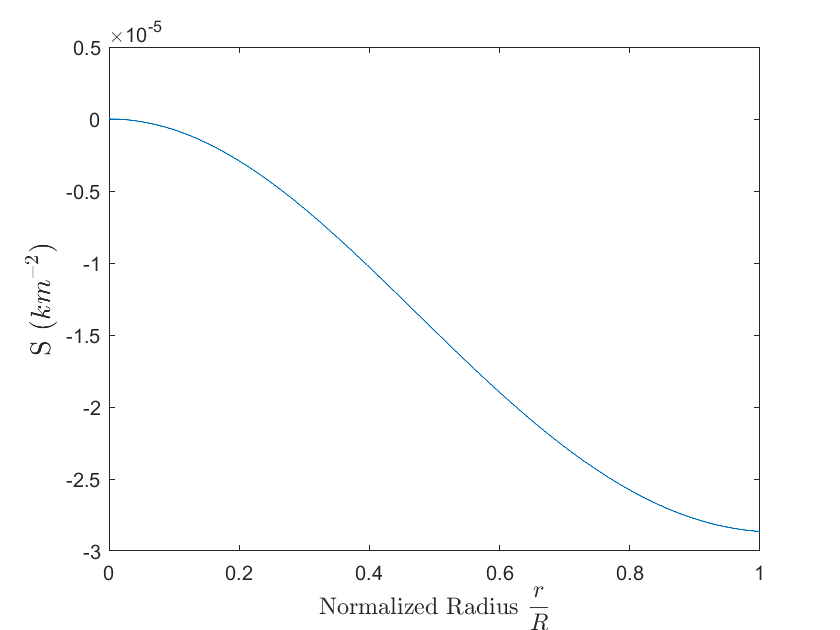}}
\subfigure[]{\includegraphics[width = .43\textwidth]{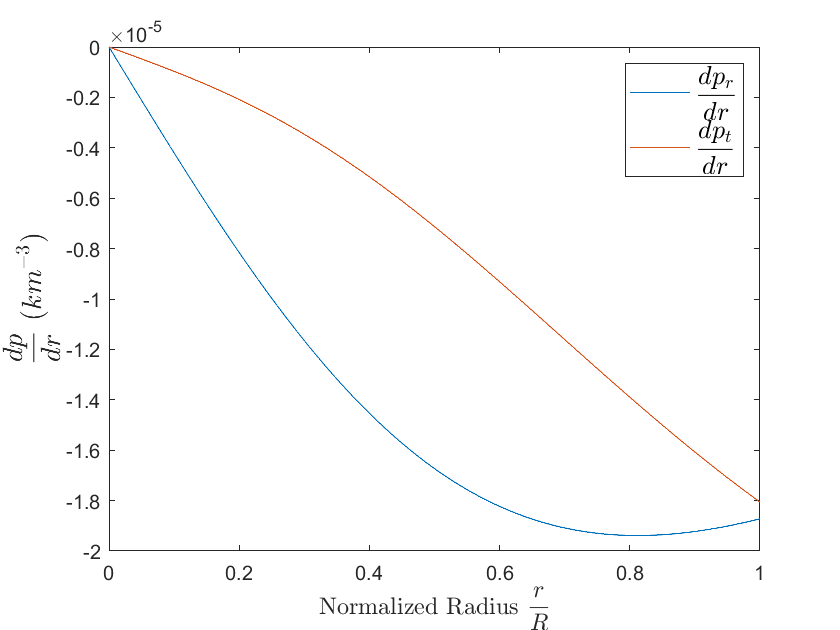}}
\subfigure[]{\includegraphics[width = .43\textwidth]{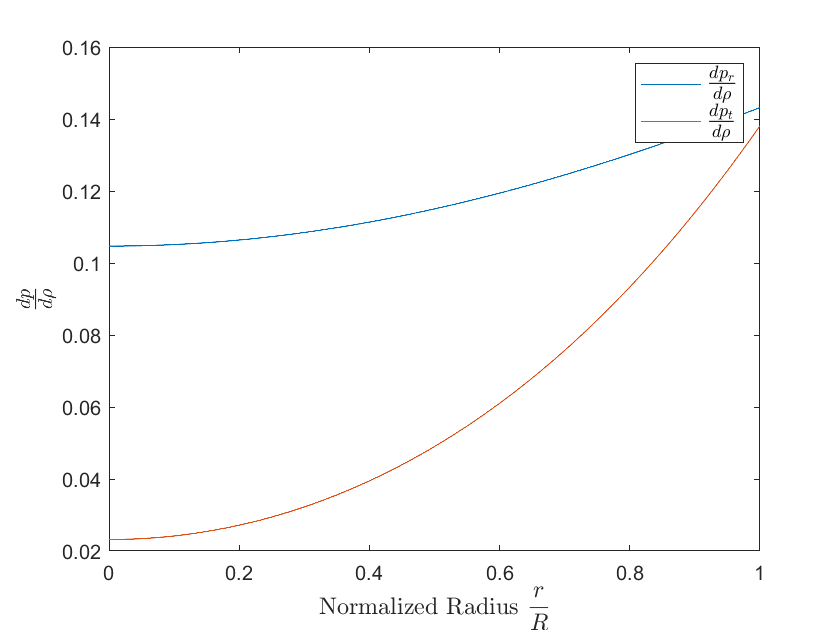}}
\subfigure[]{\includegraphics[width = .43\textwidth]{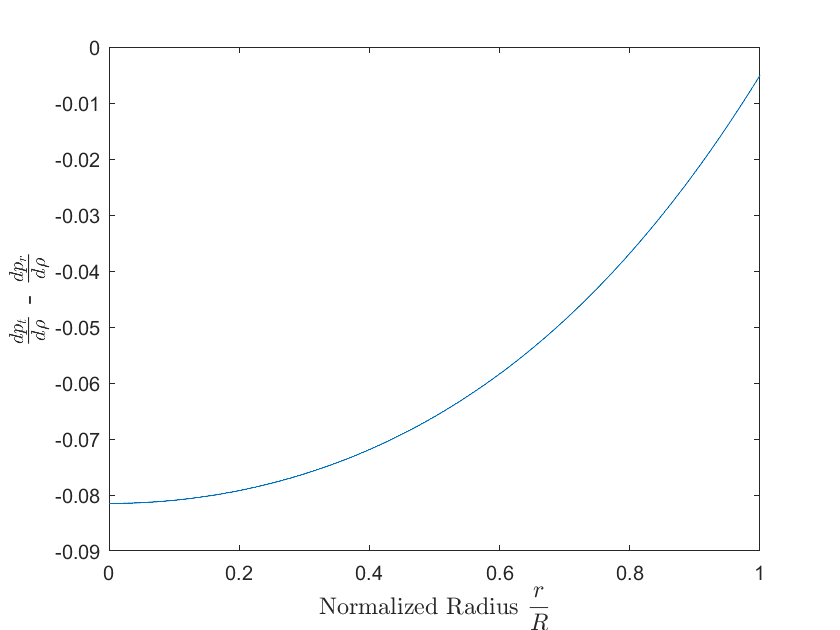}}
\subfigure[]{\includegraphics[width = .43\textwidth]{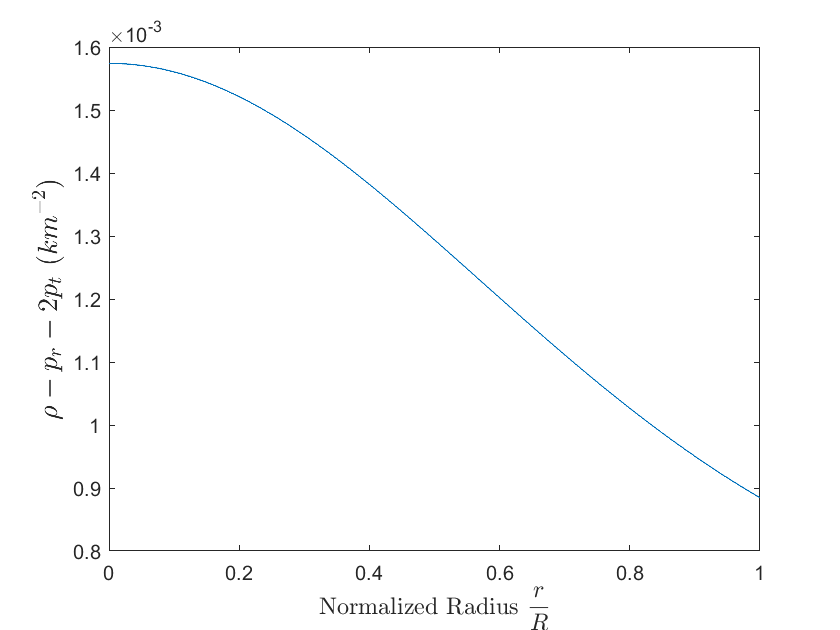}}
\subfigure[]{\includegraphics[width = .43\textwidth]{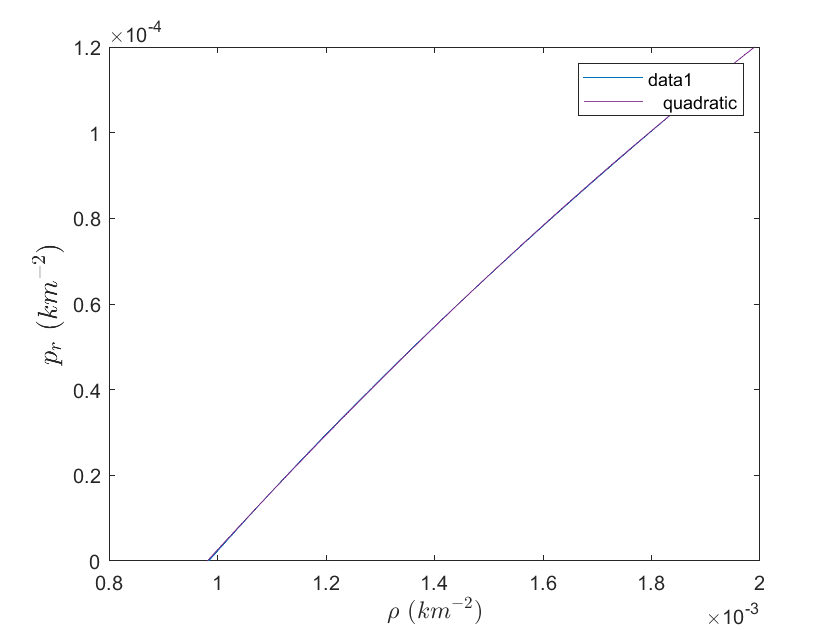}}
\caption{(a) density curve (b) pressure curves (c) anisotropy curve (d) $ \frac{dp}{dr} $ curves (e) $ \frac{dp}{d\rho} $ curves (f) $ \frac{dp_t}{d\rho} - \frac{dp_r}{d\rho} $ curve (g) $ \rho - p_r - 2p_t $ curve  (h) EOS curve for first model describing neutron star SAX J1808.4-3658(SS2) mentioned in table 1}.
\label{fig1}
\end{figure*}

We have derived a physically acceptable model of the neutron star. To predict the material composition of the stellar configuration, we need to generate an equation of the state of the stellar configuration. By generating an EOS governed by the physical laws of the system, one can parametrically relate the energy density and the radial pressure, which is useful in predicting the composition of the stellar configuration. By using equations (\ref{23}) and (\ref{25}), we have plotted the variation of the radial pressure against the energy density, as shown in figure \ref{fig1}.(h). From the plot, we find EOS of the form $ p_r = l\rho^2 + m\rho + n $.
where $l$, $m$ and $n$ are constants. By approximating EOS in figure \ref{fig1}.(h), we obtain $ l = -20 $, $ m = 0.18 $ and $ n = -0.00015 $.

\section{Second Anisotropic Model of Neutron Star in Teleparallel Gravity}\label{Sec4}
Now we develop second model for the neutron star. We consider the same metric potential as given by equation (\ref{22}), thus we get the same energy density as in equation (\ref{23}). 
\par In this new model we consider different form of radial pressure as following
\begin{equation}
\label{40}
    4\pi p_r = \frac{\frac{p_0}{R^2}\left( 1 - \frac{r^2}{R^2} \right)}{\left(1 - \frac{a r^2}{R^2}\right)^2}
\end{equation}
where $ p_0 $ is also another constant, that is the parameter of this model such that $ \frac{p_0}{R^2} $ denotes the central pressure of the neutron star. The constraint on the $ p_0 $ is that $ p_0 > 1 $, so that radial pressure remains positive for $ 0 < r < R $. The particular form of $ p_r $ is also physically reasonable because it is a monotonically decreasing function of r and satisfies the condition of vanishing radial pressure at the surface of the neutron star $ r = R $. by substituting values of equations (\ref{22}) and (\ref{40}) into (\ref{24}), we obtain
\begin{equation}
\label{41}
    c'(r) = \frac{\frac{p_0 r}{R^2}\left( 1 - \frac{r^2}{R^2} \right)}{1 - \frac{a r^2}{R^2}} + r\left( \frac{\beta_1}{4} - \frac{a}{R^2} \right) - r^3\frac{\beta_1 a}{4 R^2} 
\end{equation}
By integrating above equation, we obtain the metric as
\begin{equation}
\label{42}    
    e^{c(r)} = K \left( 1 - \frac{a r^2}{R^2} \right)^{\frac{p_0(1-a)}{2 a^2}} e^{ \frac{r^2}{2}\left(\frac{p_0}{a R^2} + \frac{\beta_1}{4} - \frac{a}{R^2} \right) -  \frac{r^4 \beta_1 a}{16 R^2}}
\end{equation}
where $ K $ is the integration constant. Now, interior space-time of the neutron star in this model is given by
\begin{equation}
\label{43}
\begin{aligned}
    ds^2 = {}& \left(K \left( 1 - \frac{a r^2}{R^2} \right)^{\frac{p_0(1-a)}{2 a^2}} e^{ \frac{r^2}{2}\left(\frac{p_0}{a R^2} + \frac{\beta_1}{4} - \frac{a}{R^2} \right) -  \frac{r^4 \beta_1 a}{16 R^2}}\right)dt^2 - \left( 1 - \frac{a r^2}{R^2} \right)dr^2 \\ &- r^2(d\theta^2+sin^2\theta d\phi^2)
\end{aligned}    
\end{equation}
\subsection{Constraining parameters for physically well-behaved model}
\subsubsection{Matching exterior space-time with interior space-time}
Now the space-time metric should be continuous at the boundary of the star. Thus, the star's interior metric (\ref{43}) should be matched to the exterior Schwarzschild space-time metric (\ref{29}) at $ r = R $ for this second model.
Now interior metric (\ref{43}) will be matched to exterior Schwarzschild metric at the surface of the star if 
\begin{equation}
\label{44}
    1 - a = \left(1 - \frac{2M}{R}\right)^{-1}
\end{equation}
and
\begin{equation}
\label{45}
    K (1 - a)^{\frac{p_0(1-a)}{2 a^2}} e^{ \left(\frac{p_0 - a^2}{2a} + \frac{\beta_1 R^2}{8}\left( 1 - \frac{a}{2}\right)\right)} =  1 - \frac{2M}{R}
\end{equation}
From equation (\ref{44}) and (\ref{45}) we obtain the value of $K$ for the second model as
\begin{equation}
\label{46}
    K = \frac{1}{(1 - a)^{\frac{p_0(1-a)}{2 a^2}+1}} e^{- \left(\frac{p_0 - a^2}{2a} + \frac{\beta_1 R^2}{8}\left( 1 -\frac{a}{2}\right)\right)}
\end{equation}
\subsubsection{Imposing physical conditions to the second model}
Any physically acceptable model of the compact star should obey physical conditions (i) to (iv) as described in section (3.1.2).
We satisfy condition (i) by imposing the following inequality computed from $ p_t \ge 0 $ in the second model.
\begin{multline}
\label{47}
    -64p_0 + R^4\beta_1^2 + 3a^2(4 + R^2\beta_1)^2 - a^3(4 + R^2\beta_1)^2 - a R^2\beta_1(16 + 3R^2\beta_1) \ge 0
\end{multline}
Now in the second model we already have $ \frac{d \rho}{dr}, \frac{d p_r}{dr} \le 0 $. In order to get $ \frac{d p_t}{dr} \le 0 $ we impose following condition to the second model
\begin{multline}
\label{48}
    -R^4\beta_1^2 - 2a^4R^2\beta_1(4 + R^2\beta_1) + 8p_0(16 + R^2\beta_1) + a^3(-16 + 24R^2\beta_1 \\+ 7R^4\beta_1^2) + a^2(-48 - 32R^2\beta_1 - 9R^4\beta_1^2 +  8p_0(4 + R^2\beta_1)) \\+ a(-16p_0(-4 + R^2\beta_1) + R^2\beta_1*(16 + 5R^2\beta_1)) \ge 0
\end{multline}
Now conditions (iii) and (iv) satisfy if the model parameters of the second satisfy following conditions, which we deduce by calculating $\frac{d p_r}{d\rho} $, $ \frac{d p_t}{d\rho} $ and $\frac{d p_t}{d\rho} - \frac{d p_r}{d\rho} $ at $ r = 0 $ and $ r = R $.
\begin{equation}
\label{49}
    0 \le \frac{p_0(1 - 2a)}{5a^2} \le 1 \quad\mathrm{and}\quad 0 \le \frac{p_0(1 - a)}{a^2(a-5)} \le 1
\end{equation}
\begin{equation}
\label{50}
    0 \le -\frac{(48a^2 - 128p_0 + 192a p_0 + 16p_0^2 - 16a R^2\beta_1 + 
     8p_0R^2\beta_1 + R^4\beta_1^2)}{320a^2} \le 1
\end{equation}
\begin{multline}
\label{51}
   0 \le \frac{1}{64(-5 + a)a^2}(R^4\beta_1^2 + 2a^4R^2\beta_1(4 + R^2\beta_1) - 8p_0(16 + R^2\beta_1) \\+  a^3(16 - 24R^2\beta_1 - 7R^4\beta_1^2) + a^2(48 + 32R^2\beta_1 + 9R^4\beta_1^2 \\- 8p_0(4 + R^2\beta_1)) + a(16p_0(-4 + R^2\beta_1) - R^2\beta_1(16 + 5R^2\beta_1))) \le 1
\end{multline}
\begin{equation}
\label{52}
    48a^2 - 64p_0 + 64a p_0 + 16p_0^2 - 16a R^2\beta_1 + 8p_0R^2\beta_1 + R^4\beta_1^2 > 0
\end{equation}
\begin{multline}
\label{53}
    R^4\beta_1^2 + 2a^4R^2\beta_1(4 + R^2\beta_1) - 8p_0(8 + R^2\beta_1) +  a^3(16 - 24R^2\beta_1 - 7R^4\beta_1^2) \\+ a^2(48 + 32R^2\beta_1 + 9R^4\beta_1^2 - 8p_0(4 + R^2\beta_1)) \\+ a(16p_0(-8 + R^2\beta_1) - R^2\beta_1(16 + 5R^2\beta_1)) > 0
\end{multline}

\subsection{Physical analysis}
This section analyzes the behavior of physical parameters such as energy density and two pressures at the interior of the star of the second model describing four observed neutron stars. We calculate parameters for each star such that this second model describing four neutron stars mentioned in Table 3. satisfy all bound conditions from (i) to (iv). We obtain Mass(M) and radius(R) of the neutron stars from references\cite{Lattimer_2014, Arzoumanian_2018} and calculate central density($  \rho_c $), surface density($ \rho_s $), central pressure($p_r(r=0) = p_t(r=0) = p_c$) and surface tangential pressure ($ p_t(r=R) = p_s $) of each star, shown in Table 4. It can be observed that this second model describe potentially stable neutron stars of higher central density and higher central pressure than neutron stars described by first model. The modification in radial pressure for second model leads to description of stable neutron stars of comparable higher density and higher  pressure profile. We have plotted various parameters of the star shown in figure \ref{fig2}. (a) to (h) of a neutron star RX J 1856-37 (1) shown in Table 3. described by this second model of the neutron stars in teleparallel gravity. The figures indicate that all the star's physical parameters are well-behaved and follow all conditions in equations (i) to (iv) at all interior points of the star.

\begin{table}[h]
\begin{center}
\caption{Values of the model parameters of the second model describing four observed neutron stars with $ \beta = 2$}\label{tab3}%
\begin{tabular}{@{}llllll@{}}
\toprule
Star & R & M & a & $ \beta_1\times 10^4 $ & $p_0$ \\ 
        & (km) & ($ M_{\odot} $) & & ($ km^{-2} $)\\
\midrule
RX J 1856-37 & 6 & 0.9031 & -0.805 & -11.944 & 0.21  \\ 
 
 Cen X-3 & 9.51 & 1.49 & -0.869 & -49.757 & 0.2 \\

 EXO 1785-248 & 8.99 & 1.3 & -0.751 & -37.119 & 0.2  \\
 
 PSR J1614-2230 & 13 & 1.908 & -0.77 & -18.343 & 0.21 \\
\botrule
\end{tabular}
\end{center}
\end{table}

\begin{table}[h]
\begin{center}
\caption{Values of the physical parameters of the four observed neutron stars described by the second model.}\label{tab4}%
\begin{tabular}{@{}lllllll@{}}
\toprule
 Star & R & M & $ \rho_c \times 10^3 $ & $ \rho_s \times 10^3 $ & $p_c\times 10^4$ & $p_s\times 10^4$ \\ 
        & (km) & ($ M_{\odot} $) & ($ km^{-2} $) & ($ km^{-2} $) & ($ km^{-2} $) & ($ km^{-2} $)\\
\midrule
 RX J 1856-37 & 6 & 0.9031 & 5.32 & 2.06 & 4.64 & 2.61 \\ 
 
 Cen X-3 & 9.51 & 1.49 & 2.19 & 0.75 & 1.76 & 0.73\\

 EXO 1785-248 & 8.99 & 1.3 & 2.15 & 0.83 & 1.97 & 0.65\\
 
 PSR J1614-2230 & 13 & 1.908 & 1.05 & 0.40 & 0.98 & 0.32\\
\botrule
\end{tabular}
\end{center}
\end{table}

\subsection{Equation of state}
\begin{figure*}[ht!]
\centering
\subfigure[]{\includegraphics[width = .43\textwidth]{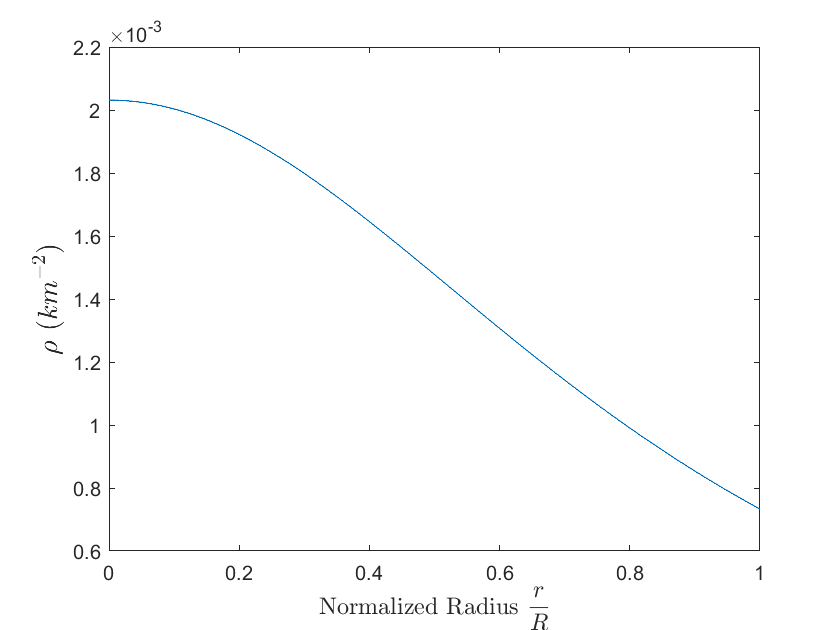}}
\subfigure[]{\includegraphics[width = .43\textwidth]{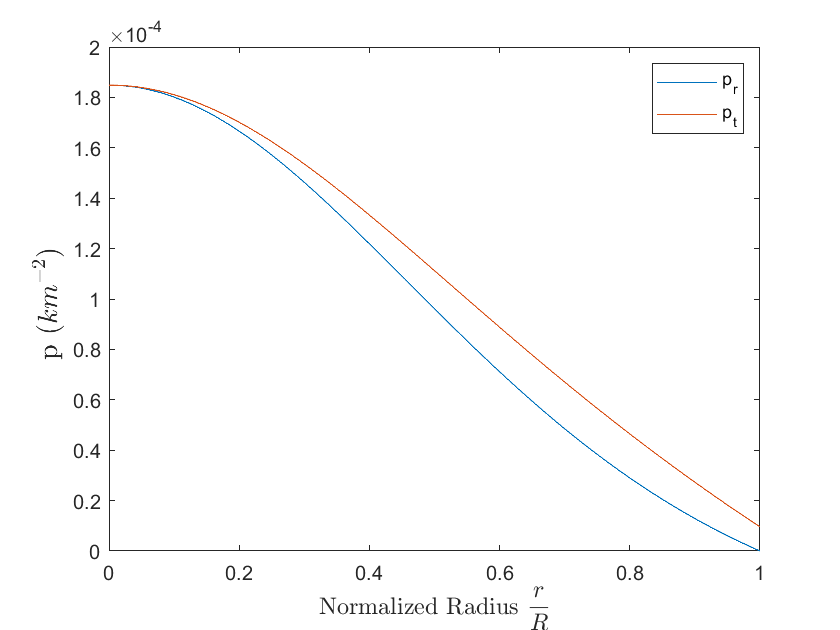}}
\subfigure[]{\includegraphics[width = .43\textwidth]{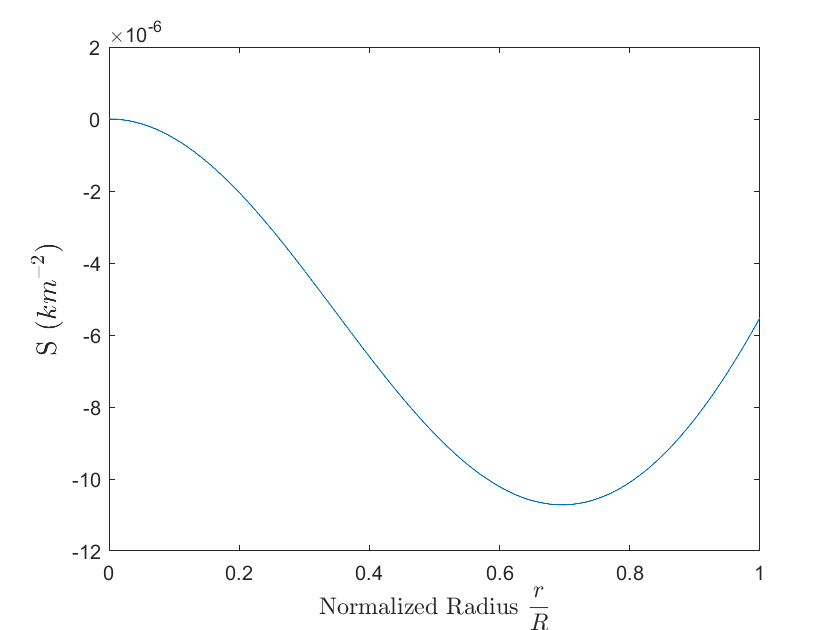}}
\subfigure[]{\includegraphics[width = .43\textwidth]{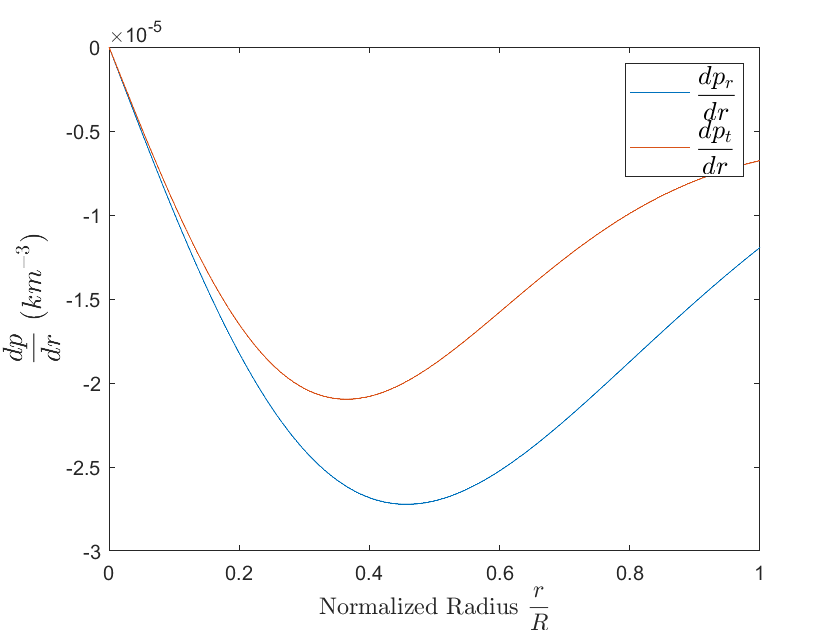}}
\subfigure[]{\includegraphics[width = .43\textwidth]{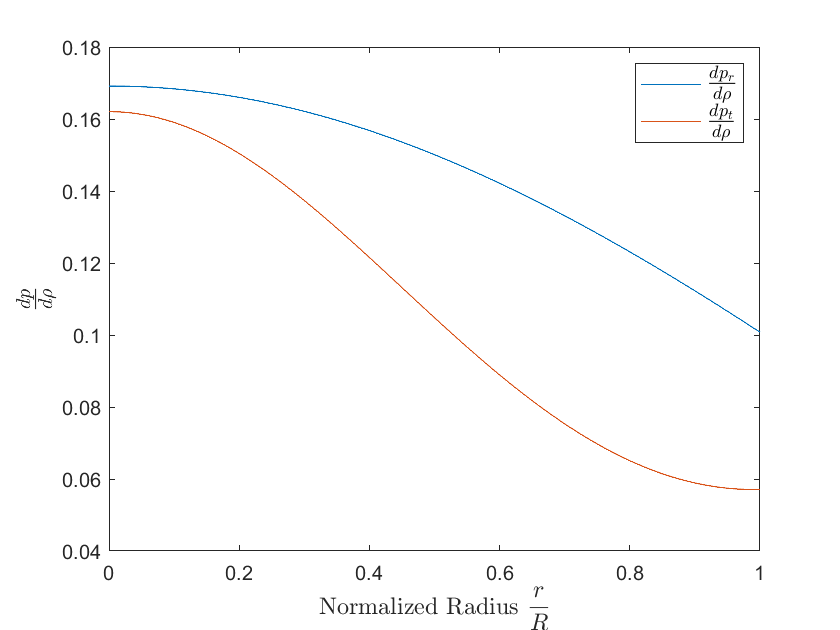}}
\subfigure[]{\includegraphics[width = .43\textwidth]{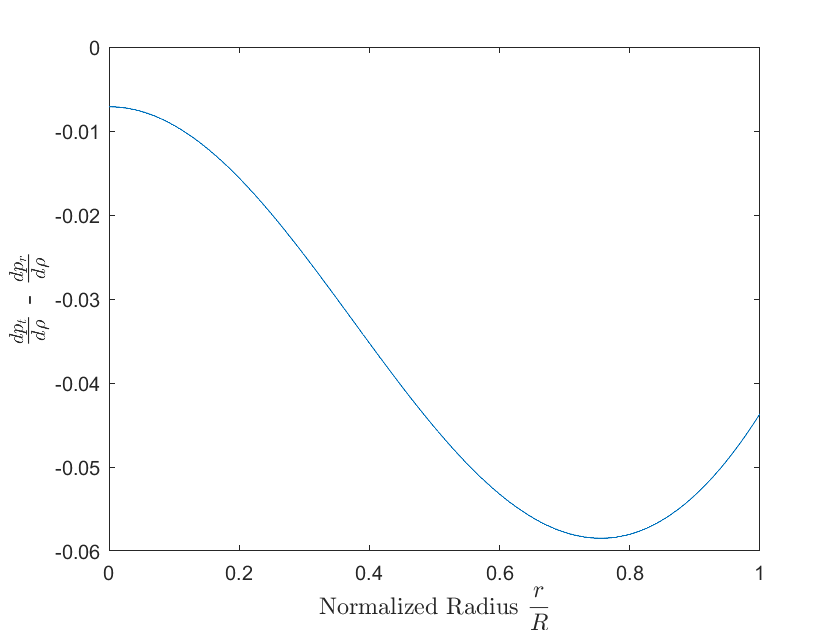}}
\subfigure[]{\includegraphics[width = .43\textwidth]{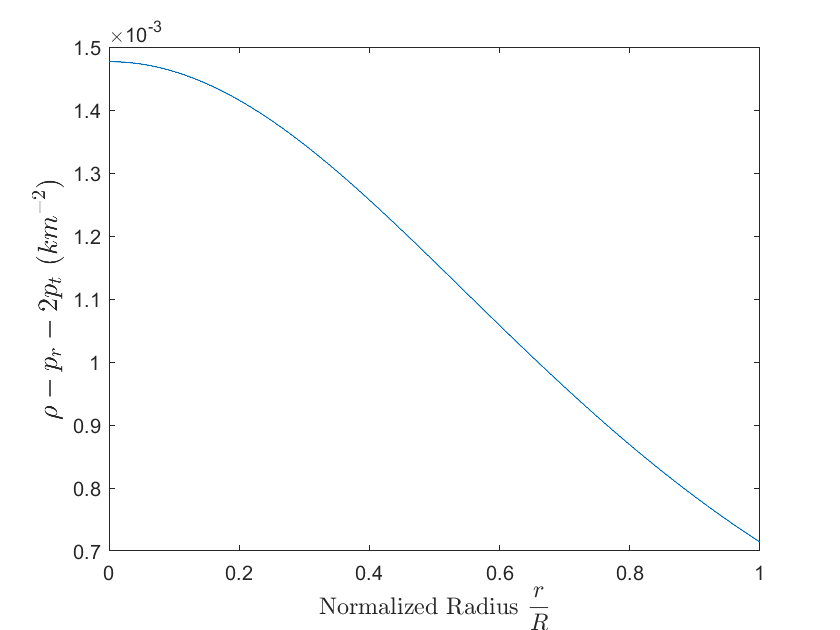}}
\subfigure[]{\includegraphics[width = .43\textwidth]{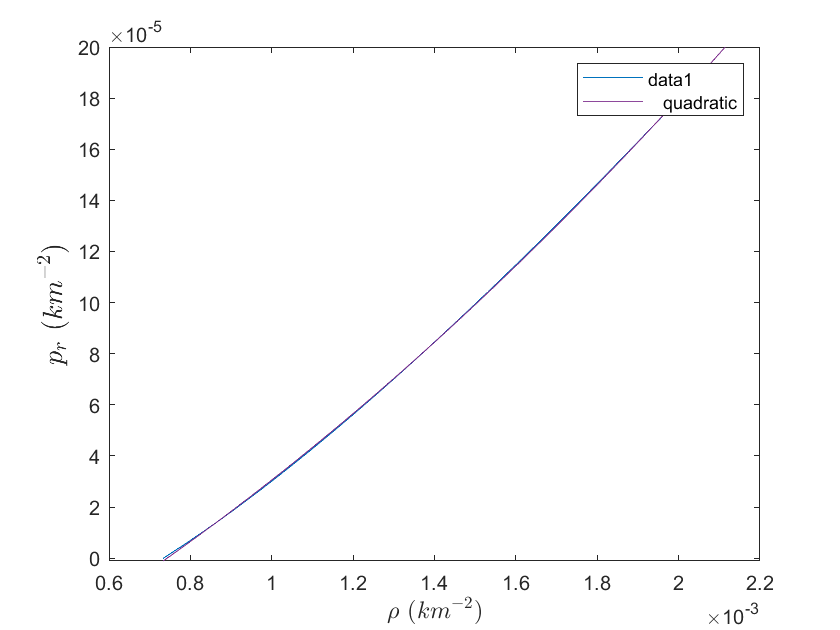}}
\caption{(a) density curve (b) pressure curves (c) anisotropy curve (d) $ \frac{dp}{dr} $ curves (e) $ \frac{dp}{d\rho} $ curves (f) $ \frac{dp_t}{d\rho} - \frac{dp_r}{d\rho} $ curve (g) $ \rho - p_r - 2p_t $ curve  (h) EOS curve for second model describing neutron star RX J 1856-37 mentioned in table 3}.
\label{fig2}
\end{figure*}

We have derived a second physically acceptable model of the neutron star. To predict the material composition of the stellar configuration, we need to generate an equation of the state of the stellar configuration for this model. By using equations (\ref{23}) and (\ref{40}), we have plotted the variation of the radial pressure against the energy density for the second model, as shown in figure \ref{fig2} (h). From the plot, we find EOS of the form $ p_r = l\rho^2 + m\rho + n $. where $l$, $m$ and $n$ are constants. By approximating EOS in figure \ref{fig2}.(h), we obtain $ l = 24 $, $ m = 0.077 $ and $ n = -7 \times 10^{-5} $.

\section{Third Anisotropic Model of Neutron Star in Teleparallel Gravity}\label{Sec5}
Now, we develop the third model for the neutron star. We consider the following form of metric potential:
\begin{equation}
\label{54}
    e^{d(r)} = \Big(1 - \frac{a r^2}{R^2}\Big)^2
\end{equation}
where a is constant and R is the radius of the neutron star. We consider $ a < 1 $, so that the metric does not become singular at any point for $ r < R $. By substituting equation (\ref{54}) in equation (\ref{17}), we obtain 
\begin{equation}
\label{55}
     4\pi \rho = \frac{\beta_{1}}{4} + \frac{\beta\Big(\frac{a^2r^2}{R^4} - \frac{2a}{R^2}\Big)}{\Big(1 - \frac{ar^2}{R^2}\Big)^2} - \frac{2a\beta}{\Big(1 - \frac{ar^2}{R^2}\Big)^3}
\end{equation}
Now from the equation (\ref{18}), we write
\begin{equation}
\label{56}
    c'(r) = (4\pi p_r)\frac{re^{d(r)}}{\beta} + \frac{e^{d(r)} - 1}{r} + \frac{\beta_1 r e^{d(r)}}{2 \beta}
\end{equation}
In order to solve the equation, we consider following form of $ 4\pi p_r $
\begin{equation}
\label{57}
    4\pi p_r = \frac{\frac{p_0}{R^2}\left( 1 - \frac{r^2}{R^2} \right)}{1 - \frac{a r^2}{R^2}}
\end{equation}
where $ p_0 $ is constant in this model also, that is the parameter of the model such that $ \frac{p_0}{R^2} $ denotes the central pressure of the neutron star. The constraint on the $ p_0 $ is that $ p_0 > 1 $, so that radial pressure remains positive for $ 0 < r < R $. The particular form of $ p_r $ is physically reasonable because it is a monotonically decreasing function of r and satisfies the condition of vanishing radial pressure at the surface of the neutron star $ r = R $. By substituting values of equations (\ref{54}) and (\ref{57}) into (\ref{56}), we obtain
\begin{equation}
\label{58}
    c'(r) = \frac{2p_0r}{\beta R^2}\Bigg(1 - \frac{r^2}{
    R^2}\Bigg)\Bigg(1 - \frac{ar^2}{R^2}\Bigg) + \frac{a^2 r^3}{R^4} - \frac{2a r}{R^2}  + \frac{r\beta_1}{2\beta}\Bigg(1 - \frac{r^2}{
    R^2}\Bigg)
\end{equation}
By integrating the above equation, we obtain the metric as
\begin{equation}
\label{59}
    e^{c(r)} = Ke^{\frac{r^6}{6}\big(\frac{2a p_0}{\beta R^6} + \frac{a^2 \beta_1}{2 \beta R^4}\big) - \frac{r^4}{4}\big(-\frac{a^2}{R^4} + \frac{2(1 + a)p_0}{\beta R^4} + \frac{a \beta_1}{\beta R^2}\big) + \frac{r^2}{2}\big(\frac{\beta_1}{2 \beta} - \frac{2 a}{R^2} + \frac{2 p_0}{\beta R^2}\big)}
\end{equation}
Now interior space-time of the neutron star in this model is given by
\begin{equation}
\label{60}
\begin{aligned}
    ds^2 = {}& \Bigg(Ke^{\frac{r^6}{6}\big(\frac{2a p_0}{\beta R^6} + \frac{a^2 \beta_1}{2 \beta R^4}\big) - \frac{r^4}{4}\big(-\frac{a^2}{R^4} + \frac{2(1 + a)p_0}{\beta R^4} + \frac{a \beta_1}{\beta R^2}\big) + \frac{r^2}{2}\big(\frac{\beta_1}{2 \beta} - \frac{2 a}{R^2} + \frac{2 p_0}{\beta R^2}\big)}\Bigg)dt^2 \\&- \left( 1 - \frac{a r^2}{R^2} \right)^2dr^2 - r^2(d\theta^2+sin^2\theta d\phi^2)
\end{aligned}    
\end{equation}
\subsection{Constraining parameters for physically well-behaved model}
\subsubsection{Matching exterior space-time with interior space-time}
Now the space-time metric should be continuous at the boundary of the star. Thus, the star's interior metric (\ref{60}) should be matched to the exterior Schwarzschild space-time metric (\ref{29}) at $ r = R $ for this third model.
Now interior metric (\ref{60}) will be matched to exterior Schwarzschild metric at the surface of the star if 
\begin{equation}
\label{61}
(1 - a)^2 = \left(1 - \frac{2M}{R}\right)^{-1}
\end{equation}
and
\begin{equation}
\label{62}
    Ke^{\frac{\beta_1 R^2}{12\beta}(a^2+3a+3) + \frac{a^2}{4} - a\big(1 + \frac{p}{6\beta}\big)}=  1 - \frac{2M}{R}
\end{equation}
From equation (\ref{61}) and (\ref{62}) we obtain the value of $K$ for the second model as
\begin{equation}
\label{63}
    K = \frac{1}{(1 - a)^2}e^{- \frac{\beta_1 R^2}{12\beta}(a^2+3a+3) + \frac{a^2}{4} - a\big(1 + \frac{p}{6\beta}\big)}
\end{equation}
For given value of $ a $, $ p_0 $, $ \beta $, $ \beta_1 $ and $ R $, value of constant $ K $ in equation (60) can be calculated from equation (63). 
\subsubsection{Imposing physical conditions to the third model}
Any physically acceptable model of the compact star should obey physical conditions (i) to (iv) as described in section (3.1.2).
We satisfy condition (i) by imposing the following inequality computed from $ p_t \ge 0 $ in the third model.
\begin{multline}
\label{64}
    32 \beta p_0 - \beta_1^2 R^4 + 5a \beta_1^2 R^4 - 5 a^4 (2 \beta + \beta_1 R^2)^2 + 
 a^5 (2 \beta + \beta_1 R^2)^2\\ + 16 a \beta (-4 p_0 + \beta_1 R^2) + 2 a^3 (24 \beta^2 + 22 \beta \beta_1 R^2 + 5 \beta_1^2 R^4) \\- 2 a^2 (24 \beta^2 - 16 \beta p_0 + 22 \beta \beta_1^2 R^4) \le 0
\end{multline}
 Now in the third model we already have $ \frac{d \rho}{dr}, \frac{d p_r}{dr} \le 0 $. In order to get $ \frac{d p_t}{dr} \le 0 $ we impose following condition to the third model
\begin{multline}
\label{65}
-64 \beta p_0 + \beta_1 R^2 (-8 p_0 + \beta_1 R^2) - 8 a^5 (2 \beta + \beta_1 R^2)(\beta - p_0 + 2 \beta_1 R^2) + \\ 8 a^3 (2 p_0 - \beta_1 R^2) (11 \beta + 5 \beta_1 R^2) + 8 a (24 \beta p_0 - 2 \beta \beta_1 R^2 + 5 \beta_1 p_0 R^2 - \beta_1^2 R^4) \\+ a^6 (4 \beta^2 + 8 \beta \beta_1 R^2 + 3 \beta_1^2 R^4) + a^2 (48 \beta^2 - 240 \beta p_0 + 56 \beta \beta_1 R^2 - 80 \beta_1 p_0 R^2 \\+ 25 \beta_1^2 R^4) + a^4 (12 \beta^2 - 80 \beta (p_0 - \beta_1 R^2) + 5\beta_1 R^2 (-8 p_0 + 7 \beta_1 R^2)) \ge 0
\end{multline}
Now conditions (iii) and (iv) satisfy if the model parameters of the third satisfy following conditions, which we deduce by calculating $\frac{d p_r}{d\rho} $, $ \frac{d p_t}{d\rho} $ and $\frac{d p_t}{d\rho} - \frac{d p_r}{d\rho} $ at $ r = 0 $ and $ r = R $.
\begin{equation}
\label{66}
0 \le -\frac{2 p_0(a - 1)}{15\beta a^2} \le 1 \quad\mathrm{and}\quad 0 \le -\frac{2 p_0(a - 1)^3}{\beta a^2 (a^2 - 4a +15)} \le 1
\end{equation}\\
\begin{equation}
\label{67}
    0 \le -\frac{48 a^2 \beta^2  - 16 a \beta \beta_1 R^{2} +
  16 p_0^2  + \beta_1^2 R^4 + 8 p_0 (-8 \beta + \beta_1 R^2)}{240 a^2 \beta^2 } \le 1
\end{equation}
\begin{multline}
\label{68}
   0 \le \frac{1}{16 a^2 (15 - 4 a + a^2) \beta^2}(64 \beta p_0 + \beta_1 R^2 (8 p_0 - \beta_1 R^2) + 8 a^5 (2 \beta + \beta_1 R^2) \\ \times (\beta - p_0 + 2 \beta_1 R^2) - 8 a^3 (2 p_0 - \beta_1 R^2)(11 \beta + 5 \beta_1 R^2) - 8 a (24 \beta p_0 \\- 2 \beta \beta_1 R^2 + 5 \beta_1 p_0 R^2 - \beta_1^2 R^4) - a^6 (4 \beta^2 + 8 \beta \beta_1 R^2 + 3 \beta_1^2 R^4) \\+ a^2 (-48 \beta^2 + 8 \beta (30 p_0 - 7 \beta_1 R^2) + 5 \beta_1 R^2 (16 p_0 - 5 \beta_1 R^2)) \\+ a^4 (-12 \beta^2 + 5 \beta_1 R^2 (8 p_0 - 7 \beta_1 R^2) + 80 \beta (p_0 - \beta_1 R^2))) \le 1
\end{multline}
\begin{equation}
\label{69}
48 a^2 \beta^2 - 32 \beta p - 16 a \beta (2 p + \beta_1 R^2) + (4 p + \beta_1 R^2)^2 < 0
\end{equation}
\begin{multline}
\label{70}
   \frac{1}{16 a^2 (a^2 - 4a + 15 ) \beta^2}(32 \beta p + \beta_1 R^2 (8 p - \beta_1 R^2) + 
   8 a^5 (2 \beta + \beta_1 R^2)\\ \times (\beta - p + 2 \beta_1 R^2) - 
   8 a^3 (18 \beta p - 11 \beta \beta_1 R^2 + 10 \beta_1 p R^2 - 5 \beta_1^2 R^4) \\- 
   8 a (12 \beta p - 2 \beta \beta_1 R^2 + 5 \beta_1 p R^2 - \beta_1^2 R^4) - 
   a^6 (4 \beta^2 + 8 \beta \beta_1 R^2 + 3 \beta_1^2 R^4) \\+ 
   a^2 (-48 \beta^2 + 8 \beta (18 p - 7 \beta_1 R^2) + 
      5 \beta_1 R^2 (16 p - 5 \beta_1 R^2)) \\+ 
   a^4 (-12 \beta^2 + 5 \beta_1 R^2 (8 p - 7 \beta_1 R^2) + 80 \beta (p - \beta_1 R^2))) > 0
\end{multline}

\subsection{Physical analysis}
This section analyzes the behavior of physical parameters such as energy density and two pressures at the interior of the star of the third model describing three observed neutron stars. We calculate parameters for each star such that this third model describing three neutron stars mentioned in Table 5. satisfy all bound conditions from (i) to (iv). We obtain Mass(M) and radius(R) of the neutron stars from references\cite{Steiner_2015, _zel_2016} and calculate central density($  \rho_c $), surface density($ \rho_s $), central pressure($p_r(r=0) = p_t(r=0) = p_c$) and surface tangential pressure ($ p_t(r=R) = p_s $) of each star, shown in Table 6. From Table 6. we can observe that this third model describes stable neutron stars of low mass and higher outer radius comparable to neutron stars described by first two models. Neutron stars described by the third model has low density and low pressure profile comparable to neutron stars described by first two models. The particular modification in metric potential and chosen model parameters for third model leads to description of stable neutron stars of lower density and lower pressure profile comparable to both the first two models. We have plotted various parameters of the star shown in figure \ref{fig3}. (a) to (h) of a neutron star PSR J0737-3039A (2) shown in Table 5. described by this third model of the neutron stars in teleparallel gravity. The figures indicate that all the star's physical parameters are well-behaved and follow all conditions in equations (i) to (iv) at all interior points of the star.

\begin{table}[h]
\begin{center}
\caption{Values of the model parameters of the third model describing three neutron stars with $ \beta_1 = 0 $.}\label{tab5}%
\begin{tabular}{@{}llllll@{}}
\toprule
Star & R & M & a & $ \beta $ & $p_0$ \\ 
        & (km) & ($ M_{\odot} $) \\
\midrule
 HER X-1 & 10.7655 & 0.85 & -0.2 & 3 & 0.101  \\ 
 
 PSR J0737-3039A & 14.68 & 1.4 & -0.181 & 3.2 & 0.09 \\

 PSR J0437-4715 & 13.8 & 1.4 & -0.194 & 3 & 0.095  \\
\botrule
\end{tabular}
\end{center}
\end{table}

\begin{table}[h]
\begin{center}
\caption{Values of the model parameters of the first model describing five observed neutron stars with $ \beta = 2 $}\label{tab6}%
\begin{tabular}{@{}lllllll@{}}
\toprule
Star & R & M & $ \rho_c \times 10^3 $ & $ \rho_s \times 10^3 $ & $p_c\times 10^4$ & $p_s\times 10^4$ \\ 
        & (km) & ($ M_{\odot} $) & ($ km^{-2} $) & ($ km^{-2} $) & ($ km^{-2} $) & ($ km^{-2} $)\\
\midrule
 HER X-1 & 10.7655 & 0.85 & 1.24 & 0.79 & 0.70 & 0.30 \\ 
 
 PSR J0737-3039A & 14.68 & 1.4 & 0.64 & 0.42 & 0.33 & 0.14\\

 PSR J0437-4715 & 13.8 & 1.4 & 0.73 & 0.47 & 0.40 & 0.17\\
\botrule
\end{tabular}
\end{center}
\end{table}

\subsection{Equation of state}
\begin{figure*}[ht!]
\centering
\subfigure[]{\includegraphics[width = .43\textwidth]{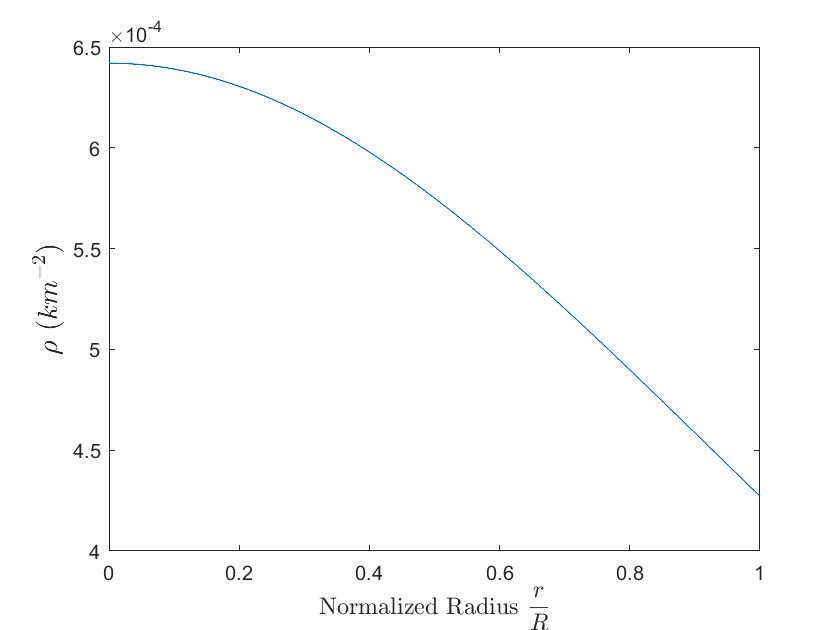}}
\subfigure[]{\includegraphics[width = .43\textwidth]{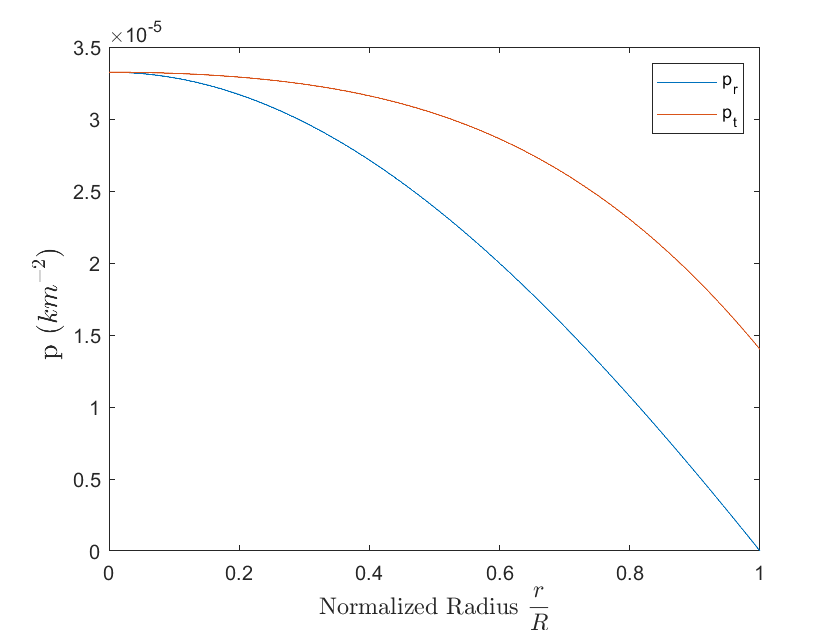}}
\subfigure[]{\includegraphics[width = .43\textwidth]{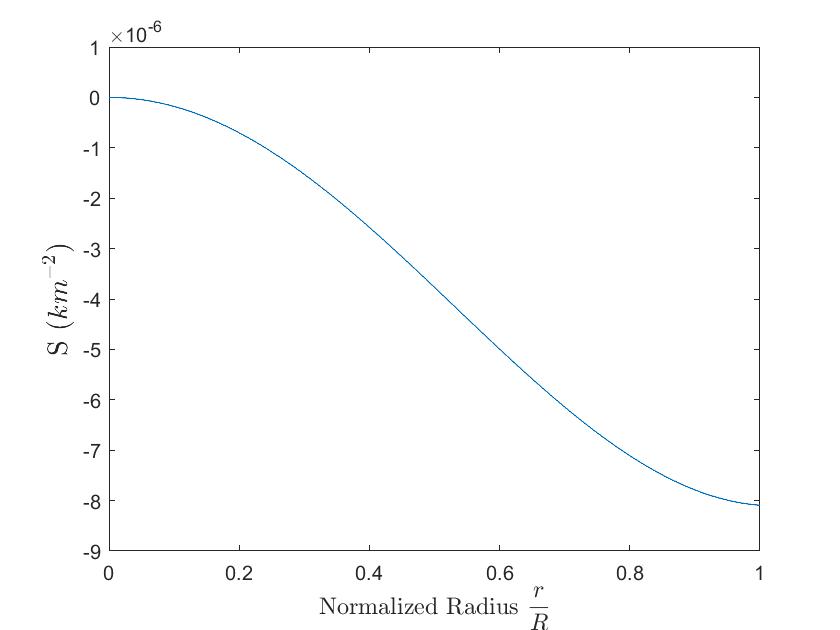}}
\subfigure[]{\includegraphics[width = .43\textwidth]{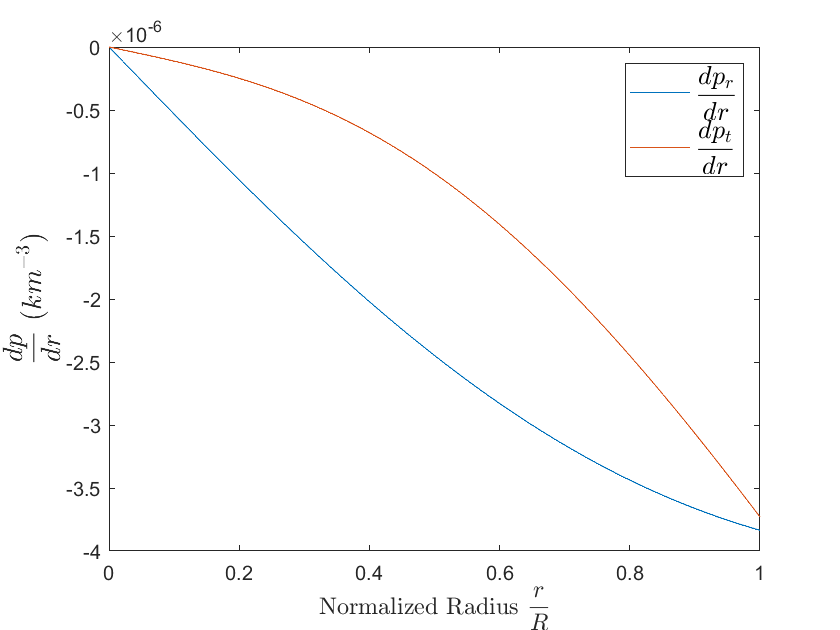}}
\subfigure[]{\includegraphics[width = .43\textwidth]{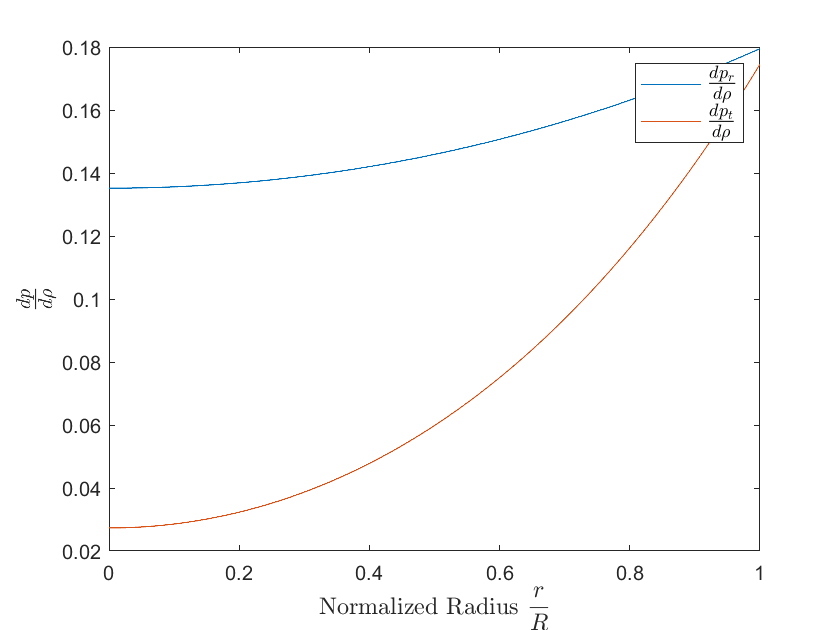}}
\subfigure[]{\includegraphics[width = .43\textwidth]{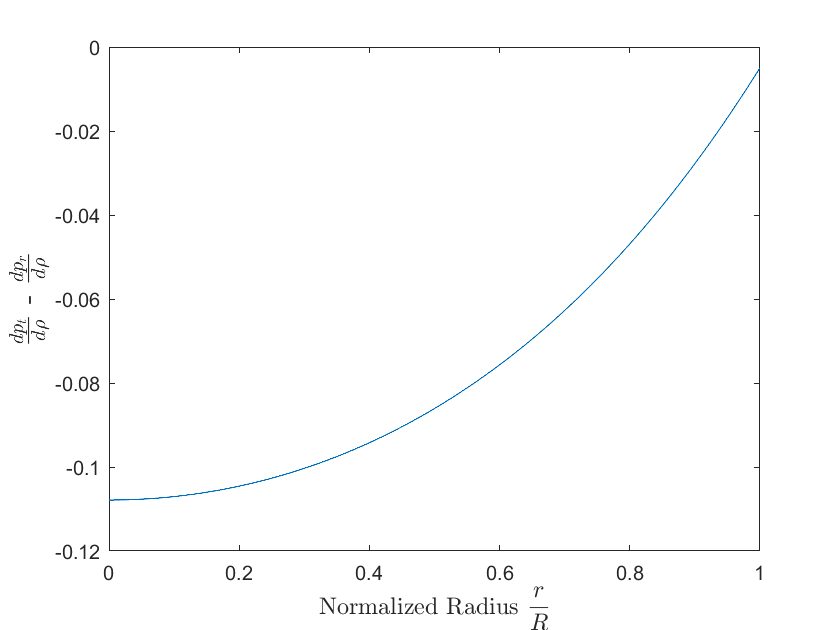}}
\subfigure[]{\includegraphics[width = .43\textwidth]{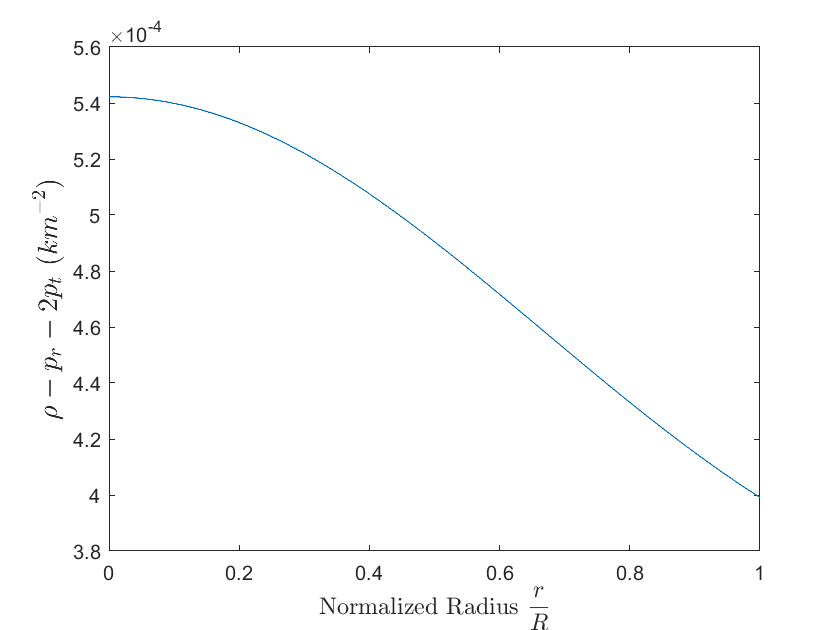}}
\subfigure[]{\includegraphics[width = .43\textwidth]{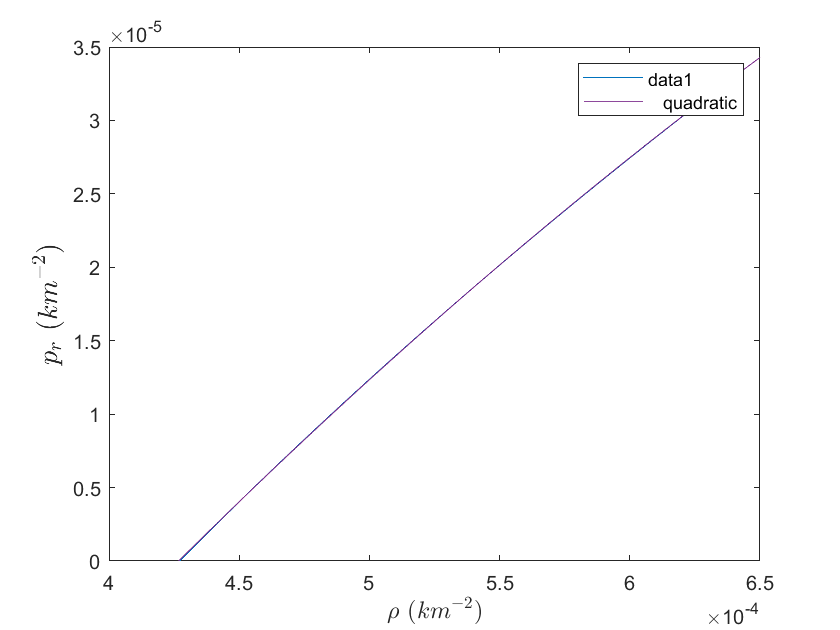}}
\caption{(a) density curve (b) pressure curves (c) anisotropy curve (d) $ \frac{dp}{dr} $ curves (e) $ \frac{dp}{d\rho} $ curves (f) $ \frac{dp_t}{d\rho} - \frac{dp_r}{d\rho} $ curve (g) $ \rho - p_r - 2p_t $ curve  (h) EOS curve for third model describing neutron star J0737-3039A mentioned in table 5}.
\label{fig3}
\end{figure*}

We have derived a third physically acceptable model of the neutron star. To predict the material composition of the stellar configuration, we need to generate an equation of the state of the stellar configuration for this model. By using equations (\ref{55}) and (\ref{57}), we have plotted the variation of the radial pressure against the energy density for the third model, as shown in figure \ref{fig3} (h). From the plot, we find EOS of the form $ p_r = l\rho^2 + m\rho + n $. where $l$, $m$ and $n$ are constants. By approximating EOS in figure \ref{fig3}.(h), we obtain $ l = -99 $, $ m = 0.26 $ and $ n = -9.3 \times 10^{-5} $.

\section{Discussion}\label{Sec6}
A new class of solutions for three analytical models describing anisotropic stellar structures for various observed neutron stars has been developed using the theory of teleparallel gravity. We have briefly reviewed the idea of tetrad field and formulation of field equations of teleparallel gravity. Then we solved equations of teleparallel gravity for spherically symmetric anisotropic matter distribution for some physically reliable metric potentials and radial pressures. Thus, we developed three analytical models describing anisotropic stellar structures of neutron stars. We calculated various model parameters by imposing physical bound conditions on them so that models become physically acceptable. We calculated model parameters and star's physical parameters for various observed neutron stars described by three models developed in this paper. We observed that particular modifications in metric potential and radial pressure leads to description of stable neutron stars ranging from low density and pressure to high density and pressure as observed in physical analysis of each model. We have shown that models with very good approximations, admits quadratic equations of state, which is very useful in determining compositions of stellar structures of particular neutron stars as well as strange stars and also very useful in modelling such realistic neutron stars. It is also worth to note that, linear form of $ f(T) $ is responsible for quadratic behaviour of EOS. The non linear forms of $ f(T) $ leads to high deviation of EOS from quadratic behaviour. Thus, the authors choose to work with linear EOS. We have also plotted graphs of various parameters of observed neutron stars described by each model developed in this paper. Graphs shows that each model is physically reliable and realistic in all aspects.  The research work presented in this paper provides understanding of how modification in model parameters can lead to description of various neutron stars described by teleparallel gravity. This work can be also utilized to investigate the reason for anisotropy in neutron stars due to pion condensation, hyperon interaction and electromagnetic forces in teleparallel gravity theory.

\end{document}